\documentclass[aps,twocolumn,pra,superscriptaddress,amsmath,showpacs,tightenlines]{revtex4}
\usepackage{epsfig,graphicx,times}
\usepackage{amstext}
\usepackage{amsmath}            
\usepackage{amssymb}            
\usepackage{graphicx}           
\usepackage{latexsym}
\usepackage{bm}

\usepackage[colorlinks,citecolor=blue, linkcolor=blue,hyperindex,CJKbookmarks,dvipdfm]{hyperref}

\begin{document}

\title{Strong photon antibunching of symmetric and antisymmetric modes in
weakly nonlinear photonic molecules}
\author{Xun-Wei Xu}
\affiliation{Beijing Computational Science Research Center, Beijing 100084, China}
\author{Yong Li}
\email{liyong@csrc.ac.cn}
\affiliation{Beijing Computational Science Research Center, Beijing 100084, China}
\affiliation{Synergetic Innovation Center of Quantum Information and Quantum Physics,
University of Science and Technology of China, Hefei, Anhui 230026, China}
\date{\today }

\begin{abstract}
We study the photon statistics of symmetric and antisymmetric modes in a
photonic molecule consisting of two linearly coupled nonlinear cavity modes. Our
calculations show that strong photon antibunching of both symmetric and
antisymmetric modes can be obtained even when the nonlinearity in the photonic
molecule is weak. The strong antibunching effect results from the
destructive interference between different paths for two-photon excitation.
Moreover, we find that the optimal frequency detunings for strong photon antibunching in the symmetric and antisymmetric modes are linearly dependent on the
coupling strength between the cavity modes in the photonic molecule. This
implies that the photonic molecules can be used to generate tunable single-photon sources by tuning the values of the
coupling strength between the cavity modes with weak nonlinearity.
\end{abstract}

\maketitle


\section{Introduction}

Single-photon source is one of the fundamental devices for
quantum information processing at single-photon level. In order to create a
single-photon source, Imamoglu \emph{et al}. proposed using a
high-finesse cavity containing a low density four-level atomic medium~\cite%
{ImamogluPRL97}. They found that the transmitted photons show strong
antibunching. This effect comes from the strong photon-photon interaction:
the excitation of a first photon blocks the transport of a second photon for
the nonlinear medium in the cavity, called the photon blockade effect.
Photon blockade is one of the mechanisms for creating strong antibunching
photons. In 2005 photon blockade was observed in an optical
cavity with one trapped atom~\cite{BirnbaumNat05, DayanSci08}. Subsequently, a
sequence of experimental groups observed the strong antibunching behaviors
in different systems: a quantum dot in a photonic crystal~\cite{FaraonNP08},
circuit cavity quantum electrodynamics (QED) systems~\cite%
{LangPRL11,HoffmanPRL11,LiuPRA14}.

Recently, Liew and Savona found a new mechanism in a photonic molecule consisting of two linearly coupled nonlinear
cavity modes that can give rise to strong photon
antibunching even with nonlinearities much smaller than the decay rates of the cavity modes~\cite{LiewPRL10}.
The physical explanation is that the strong photon-photon correlation was attributed to
the destructive quantum interference effect in the nonlinear photonic
molecule~\cite{BambaPRA11,CarusottoRMP13}. Based on this mechanism, many different systems
are proposed to achieve photon blockade, such as bimodal optical cavity with
a quantum dot~\cite{MajumdarPRL12,ZhangPRA14}, coupled optomechanical systems~\cite{XuJPB13,SavonaARX13}, a double quantum well embedded
in a micropillar optical cavity~\cite{KyriienkoarX14}, and coupled single-mode cavities with second- or
third-order nonlinearity~\cite{FerrettiNJP13,FlayacPRA13,GeracePRA14}.

The statistic properties of photons in the photonic molecules have already been studied in Refs.~\cite{LiewPRL10,BambaPRA11,CarusottoRMP13,MajumdarPRL12,ZhangPRA14,XuJPB13,SavonaARX13,KyriienkoarX14,FerrettiNJP13,FlayacPRA13,GeracePRA14},
focusing on the statistic properties of photons for modes located
in one of the cavities. However, as the coupling between the photonic cavities reaches the strong coupling regime, the photonic eigenmodes are the symmetric and antisymmetric modes spanning the whole system~\cite{IlchenkoOC94,NakagawaAPL05,GrudininPRL10,PengOL12,PengNP14a,ChangNP14,PengNP14b,LudwigPRL12,StannigelPRL12,KomarPRA13}. Nonclassical photon correlations for the symmetric and antisymmetric modes in two-mode optomechanics have already been studied theoretically, and it was shown that the nonlinear interactions can be significantly enhanced in the coupled optomechanics~\cite{StannigelPRL12,KomarPRA13}.

In this paper, we will investigate the photon statistics of
the symmetric and antisymmetric modes, instead of local modes, in a photonic molecule consisting of
two linearly coupled nonlinear cavities, and show that the photons of both the
symmetric and antisymmetric modes can exhibit strong antibunching effect
even with weak nonlinearity in the photonic molecule. Most importantly, different from the result given in Refs.~\cite{LiewPRL10,BambaPRA11}, we find that the optimal frequency detunings for strong photon antibunching in the symmetric and antisymmetric modes are linearly dependent on the coupling strength between the cavity modes. So we can generate tunable single-photon sources by the symmetric and antisymmetric modes in the photonic molecules with weak nonlinearity.

The paper is organized as follows: In Sec.~II, we will show the Hamiltonian
and the dynamic equation of the photonic molecule system. In
Sec.~III, the statistic properties of the photons of the symmetric and
antisymmetric modes in the photonic molecule are investigated via the
second-order correlation functions by numerical calculations. In Sec.~IV,
the optimal conditions for strong antibunching effect are obtained
analytically. Finally, we draw our conclusions in Sec.~V.

\section{Physical model}

\begin{figure}[tbp]
\includegraphics[bb=40 340 570 603, width=8.5 cm, clip]{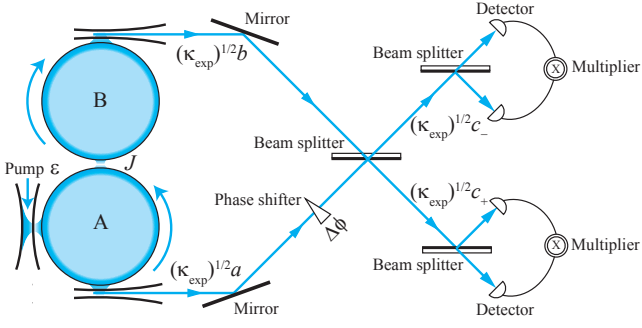}
\caption{(Color online) Schematic diagram of setup for the detection of photon antibunching effects of symmetric and antisymmetric modes in a photonic molecule consisting of two tunnel-coupled nonlinear microtoroids. $\kappa_{\rm exp}$ is the loss rate for detection.}
\label{fig1}
\end{figure}

Photonic molecule consists of two nonlinear cavity modes with coupling strength $J$. Taking the coupled microtoroids for an example, as shown in Fig.~\ref{fig1}, the coupling strength between the cavity modes in the two microtoroids depends exponentially upon the air gap~\cite{IlchenkoOC94,NakagawaAPL05}. The distance and hence the coupling between the cavity modes in the microtoroids can be controlled precisely~\cite{GrudininPRL10,PengOL12,PengNP14a,ChangNP14,PengNP14b}. The Hamiltonian for the compound system in a frame rotating at the frequency of the driving field $\omega _{d}
$ reads~\cite{LiewPRL10} ($\hbar =1$):
\begin{eqnarray}  \label{eq:1}
H &=&\Delta _{a}a^{\dag }a+\Delta _{b}b^{\dag }b-J\left( a^{\dag
}b+b^{\dag}a\right)  \notag \\
&&+Ua^{\dag }a^{\dag }aa+Ub^{\dag }b^{\dag }bb+\varepsilon \left(
a^{\dag}+a\right),
\end{eqnarray}%
where $a$ ($b$) is a bosonic operator for cavity mode A (B) with
frequency $\omega _{a}$ ($\omega _{b}$); $U$ is the Kerr nonlinear
interaction strength in each cavity. $\varepsilon $ is the Rabi frequency of the
external driving field and has been assumed to be real; $\Delta _{a}=\omega _{a}-\omega _{d}$ ($\Delta
_{b}=\omega _{b}-\omega _{d}$) is the frequency detuning between the cavity
mode and the driving field. For simplicity, we assume that the frequencies
of the two cavity modes are the same, i.e. $\Delta_{a}=\Delta _{b}=\Delta$,
then the cavity modes in the photonic molecule can be combined to form the
symmetric and antisymmetric modes by $c_{\pm }=\left( a\pm b\right) / \sqrt{2%
}$, and the Hamiltonian is transformed into
\begin{eqnarray}  \label{eq:2}
H &=&\left( \Delta-J\right) c_{+}^{\dag }c_{+}+\left( \Delta
+J\right) c_{-}^{\dag }c_{-}  \notag \\
&&+\frac{U}{2}\left( c_{+}^{\dag }c_{+}^{\dag }c_{+}c_{+}+c_{-}^{\dag
}c_{-}^{\dag }c_{-}c_{-}\right)  \notag \\
&&+\frac{U}{2}\left( c_{-}^{\dag }c_{-}^{\dag }c_{+}c_{+}+c_{+}^{\dag
}c_{+}^{\dag }c_{-}c_{-}+4c_{+}^{\dag }c_{+}c_{-}^{\dag }c_{-}\right)  \notag
\\
&&+\frac{\varepsilon }{\sqrt{2}}\left( c_{+}^{\dag }+c_{-}^{\dag }\right) +%
\frac{\varepsilon }{\sqrt{2}}\left( c_{+}+c_{-}\right).
\end{eqnarray}
There are only nonlinear couplings between the symmetric and antisymmetric modes~\cite{XiaoOE08}.

The dynamics of the system can be described by the master equation for the
density matrix $\rho $,
\begin{eqnarray}  \label{eq:3}
\frac{\partial \rho }{\partial t} &=&-i\left[ H,\rho \right]  \notag \\
&&+\frac{\kappa _{a}+\kappa _{b}}{4}\left( 2c_{+}\rho c_{+}^{\dag
}-c_{+}^{\dag }c_{+}\rho -\rho c_{+}^{\dag }c_{+}\right)  \notag \\
&&+\frac{\kappa _{a}+\kappa _{b}}{4}\left( 2c_{-}\rho c_{-}^{\dag
}-c_{-}^{\dag }c_{-}\rho -\rho c_{-}^{\dag }c_{-}\right)  \notag \\
&&+\frac{\kappa _{a}-\kappa _{b}}{4}\left( 2c_{+}\rho c_{-}^{\dag
}-c_{+}^{\dag }c_{-}\rho -\rho c_{+}^{\dag }c_{-}\right)  \notag \\
&&+\frac{\kappa _{a}-\kappa _{b}}{4}\left( 2c_{-}\rho c_{+}^{\dag
}-c_{-}^{\dag }c_{+}\rho -\rho c_{-}^{\dag }c_{+}\right),
\end{eqnarray}%
where $\kappa_{a}$ ($\kappa_{b}$) is the dissipation rate which includes the loss rate $\kappa_{\rm exp}$ (i.e. wave guide coupling) for detection. The equilibrium mean thermal photon numbers in cavity modes at optical
frequencies have been neglected. Without loss of generality, we assume that the
dissipation rates of the cavity modes are equal, i.e. $\kappa_{a}=\kappa_{b}=%
\kappa$, then the coupling terms induced by the dissipation in the master
equation (last two terms) vanish. The master equation can be solved by
expanding the density matrix over a Fock basis~\cite{LiewPRL10,VergerPRB06,XuJPB13}.

In this paper, we will focus on the statistic properties of photons for the
symmetric and antisymmetric modes in the photonic molecule, which are described by the second-order correlation functions as
\begin{equation}  \label{eq:4}
g_{\pm }^{\left( 2\right) }\left( \tau \right) =\frac{\left\langle c_{\pm
}^{\dag }\left(0\right)c_{\pm }^{\dag }\left(\tau\right) c_{\pm }\left(\tau\right) c_{\pm }\left(0\right)\right\rangle }{\left\langle c_{\pm
}^{\dag }\left(0\right)c_{\pm }\left(0\right)\right\rangle^{2} }
\end{equation}
in the steady state, where $\tau$ is the time delay between different detectors. In experiments, the statistic properties of photons for the
symmetric and antisymmetric modes can be obtained by combining the two output fields from cavity modes A and B through a 50/50 beam splitter, and detecting the statistic properties of photons for the
symmetric and antisymmetric modes individually by the Hanbury Brown-Twiss experiment~\cite{Bachor2004}, as shown in Fig.~\ref{fig1}.
In theory, we can solve the master equation numerically to get the density matrix $\rho $ within a
truncated Fock space, then the second-order correlation functions for the
symmetric and antisymmetric modes are obtained.

\section{Numerical results}

\begin{figure}[tbp]
\includegraphics[bb=71 235 498 608, width=4.2 cm, clip]{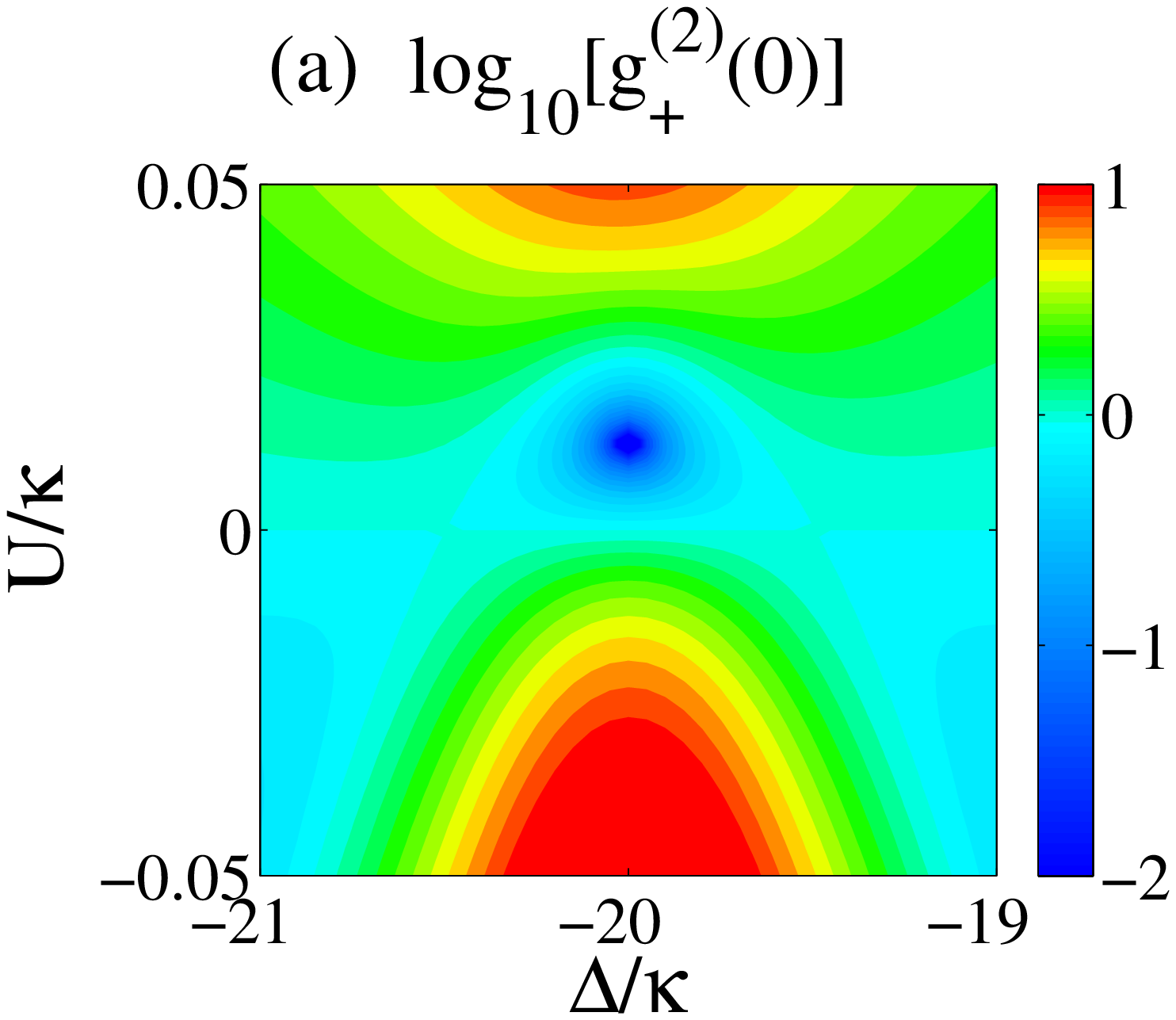}
\includegraphics[bb=71 235 498 608, width=4.2 cm, clip]{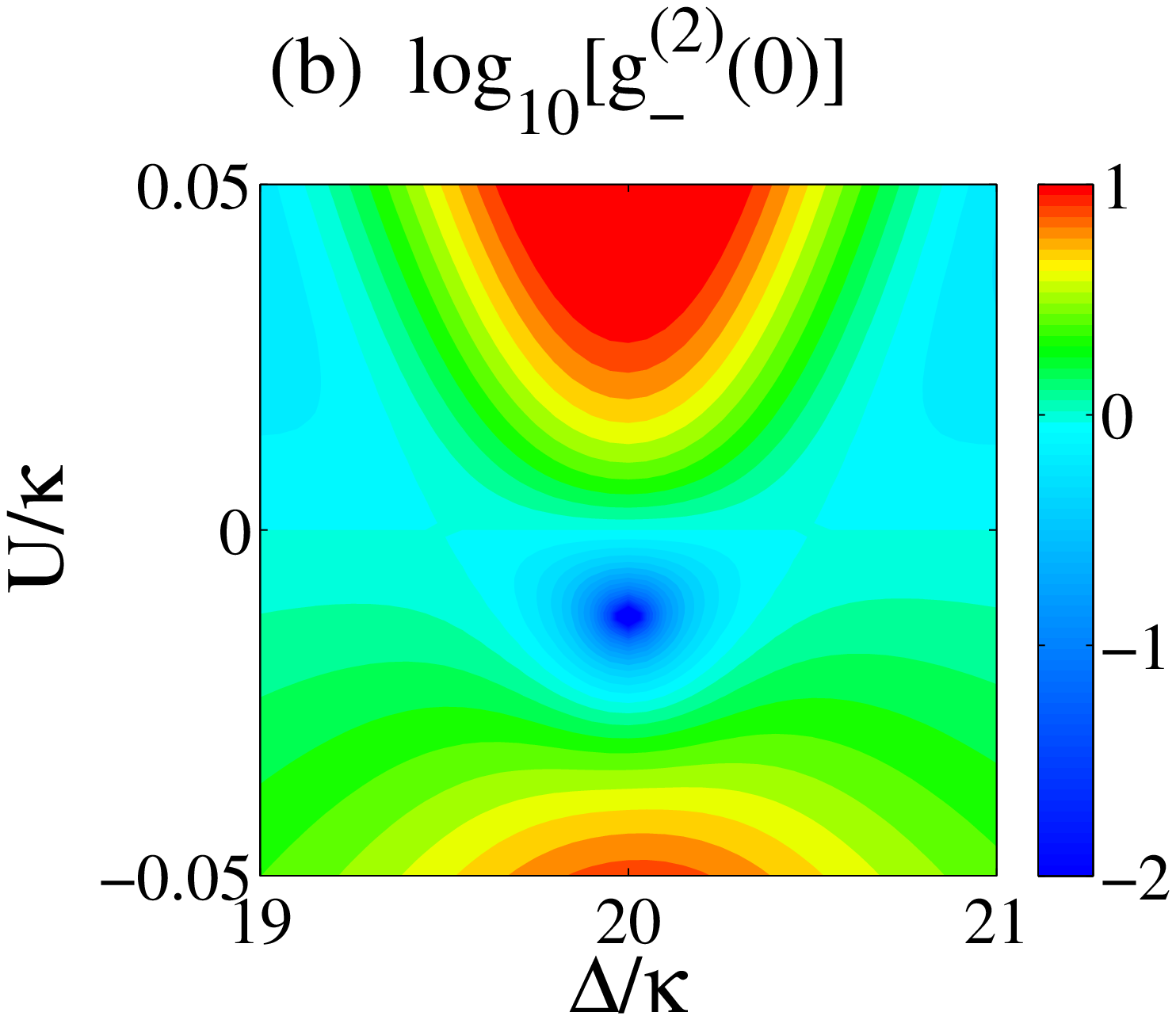}
\caption{(Color online) Logarithmic plot (of base $10$) of the equal-time
second-order correlation functions $g_{\pm}^{\left( 2\right) }\left( 0\right)$
as functions of the detuning $\Delta$ and the nonlinear interaction strength $U/\kappa $ for
the coupling strength $J=20\kappa$ and Rabi frequency $\varepsilon=0.01\kappa$.}
\label{fig2}
\end{figure}

\begin{figure}[tbp]
\includegraphics[bb=72 11 378 277, width=4.2 cm, clip]{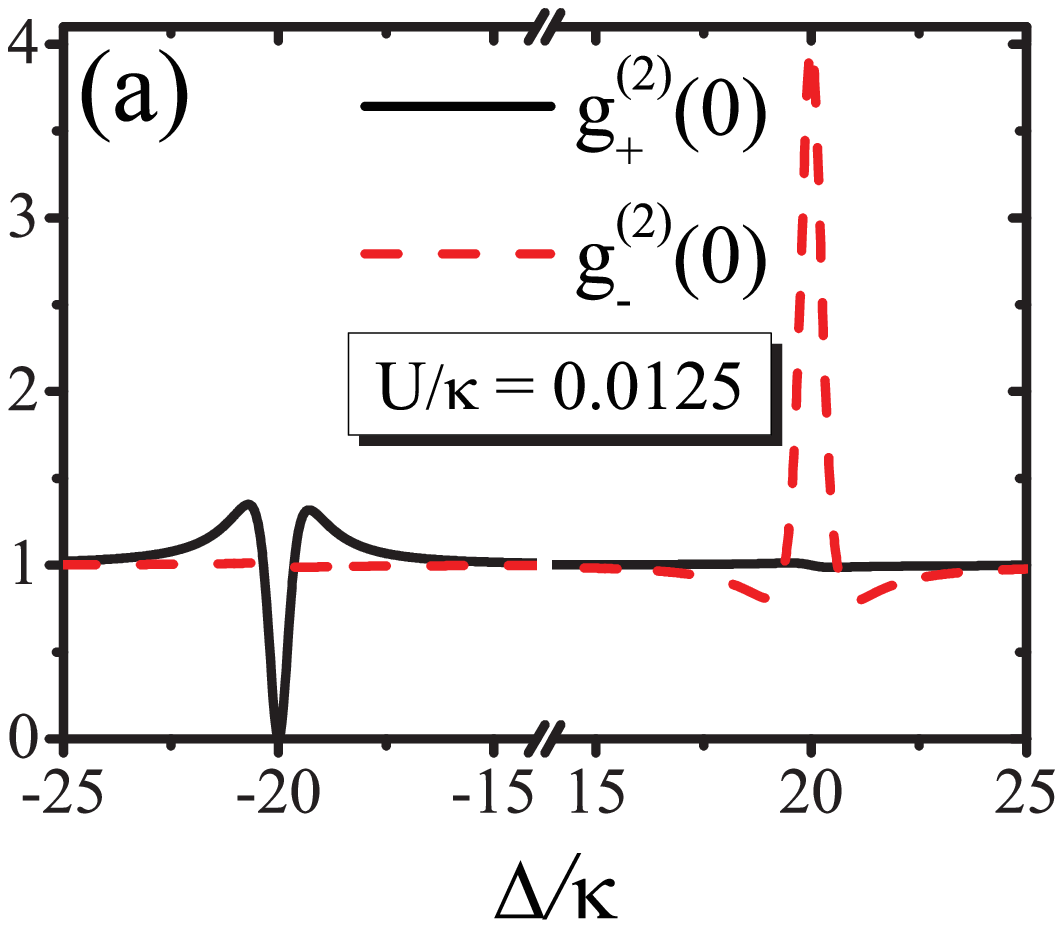}
\includegraphics[bb=72 11 378 277, width=4.2 cm, clip]{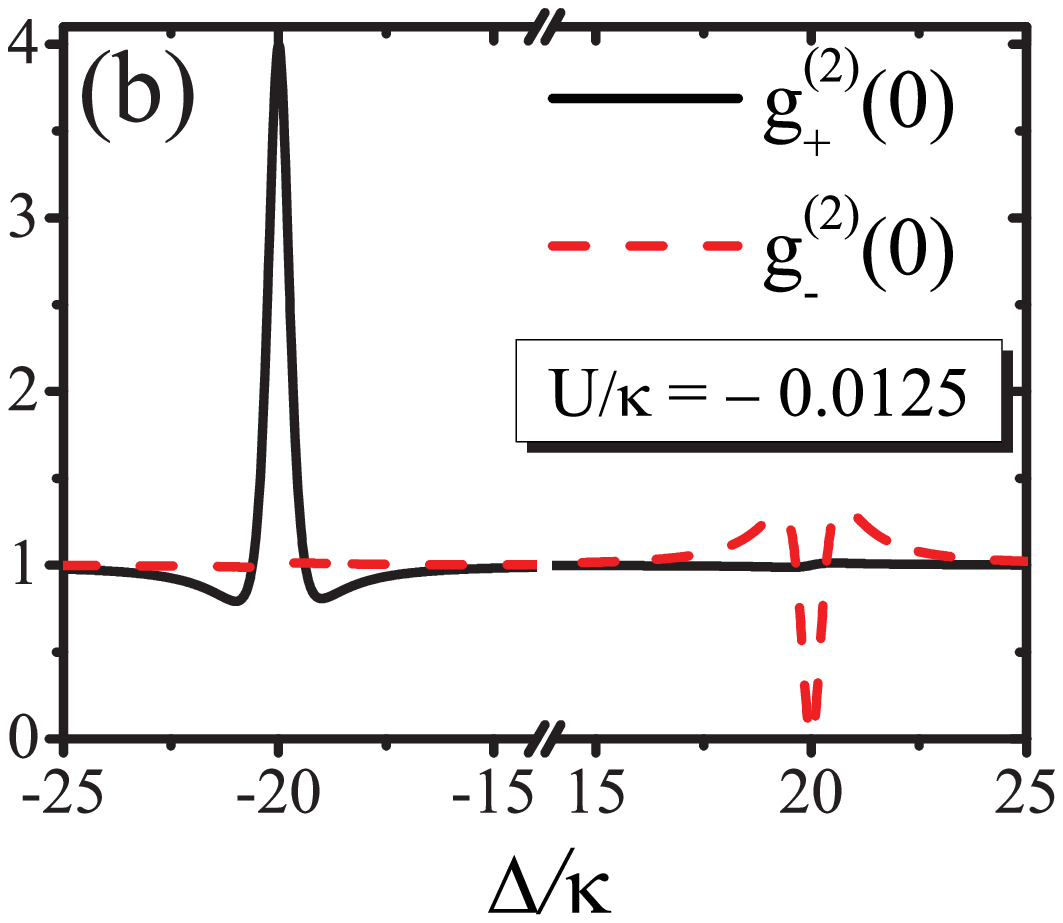}
\caption{(Color online) The equal-time second-order correlation functions $%
g_{\pm }^{\left( 2\right) }\left( 0\right)$ plotted as functions of the
detuning $\Delta /\kappa$ for nonlinear interaction strength (a)
$U/\protect\kappa=0.0125$ and (b) $U/\protect\kappa=-0.0125$. The parameters
are $J=20\protect\kappa$ and $\protect\varepsilon=0.01\protect\kappa$.}
\label{fig3}
\end{figure}

In this section, the second-order correlation functions $
g_{\pm}^{\left( 2\right) }\left( \tau \right)$ will be plotted as functions of
various parameters by solving the master equation numerically within a
truncated Fock space. We assume that the external driving field are weak, with
Rabi frequency $\varepsilon=0.01\kappa$. Such a weak
driving condition is a necessary condition for photon blockade~\cite{LiuPRA14}
and small truncated Fock space (five photons are retained in the following numerical
calculations). For convenience, we normalize all the parameters to the
dissipation rate of the cavity modes $\kappa$.

In order to find the optimal conditions for strong antibunching numerically, we show the logarithmic plot of
the equal-time second-order correlation functions $g_{\pm}^{\left( 2\right)
}\left( 0\right)$ as functions of the detuning $\Delta/\kappa$ and the nonlinear interaction strength $U/\kappa $ for
the coupling strength $J=20\kappa$ in Fig.~\ref{fig2}.
We note that there is a dip regime for $g_{+}^{\left( 2\right)
}\left( 0\right) \ll 1$ around the point $\Delta=-20\kappa$, $U=0.0125\kappa$, corresponding to strong antibunching in the symmetric mode.
Similarly, there is a dip regime for $g_{-}^{\left( 2\right)
}\left( 0\right) \ll 1$ around the point $\Delta=20\kappa$, $U=-0.0125\kappa$, corresponding to strong antibunching in the antisymmetric mode.

In Fig.~\ref{fig3}, we show the equal-time second-order correlation
functions $g_{\pm }^{\left( 2\right) }\left( 0\right) $ as functions of the
detuning $\Delta/\kappa $. For the nonlinear interaction strength $
U/\kappa =0.0125$ and coupling strength $J/\kappa =20$, $
g_{+}^{\left( 2\right) }\left( 0\right)\ll 1$ at $\Delta
=-J=-20\kappa$, while $g_{-}^{\left( 2\right) }\left( 0\right) >1$ at $\Delta=J=20\kappa$. On the contrary, for the nonlinear interaction
strength $U/\kappa =-0.0125$, $g_{-}^{\left( 2\right) }\left( 0\right) \ll 1$
around the point $\Delta=J=20\kappa$, while $g_{+}^{\left( 2\right) }\left(
0\right) >1$ around the point $\Delta=-J=-20\kappa$. These indicate that, the
photons for the symmetric (antisymetric) mode exhibit strong antibunching effect as the
detuning $\Delta=-J$ ($\Delta=J$) with weak nonlinear
interaction strength $U/\kappa =0.0125$ ($U/\kappa =-0.0125$).

\begin{figure}[tbp]
\includegraphics[bb=51 2 366 280, width=5 cm, clip]{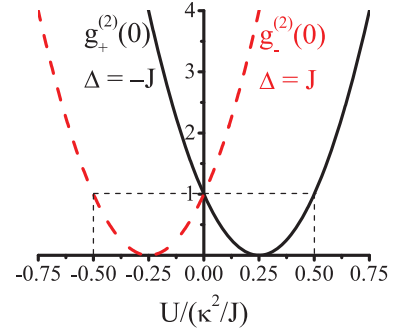}
\caption{(Color online) The equal-time second-order correlation functions $%
g_{\pm }^{\left( 2\right)}\left( 0\right)$ as functions of nonlinear
interaction strength $U$ normalized to $\protect\kappa^{2}/J$ with Rabi frequency $\varepsilon=0.01\kappa$. (Black solid line) $g_{+}^{\left( 2\right)}\left( 0\right)$ for $\Delta =-J =-20%
\protect\kappa$; (Red dash line) $g_{-}^{\left( 2\right)}\left( 0\right)$
for $\Delta =J =20\protect\kappa$.}
\label{fig4}
\end{figure}

The equal-time second-order correlation functions $g_{\pm }^{\left( 2\right)
}\left( 0\right)$ as functions of nonlinear interaction strength $U$
normalized to $\kappa^{2}/J$ is shown in Fig.~\ref{fig4}.
This plot show that the photons for the symmetric mode exhibit antibunching
as nonlinear interaction strength $0<U/(\kappa^{2}/J)<1/2$ with $\Delta
=-J$ and reach optimal strong antibunching at $U/(\kappa^{2}/J)=1/4$;
the photons for the antisymmetric mode exhibit antibunching as
nonlinear interaction strength $-1/2<U/(\kappa^{2}/J)<0$ with the detuning $%
\Delta=J$ and reach optimal strong antibunching at $U/(\kappa^{2}/J)=-1/4$.

\begin{figure}[tbp]
\includegraphics[bb=70 235 494 608, width=6 cm, clip]{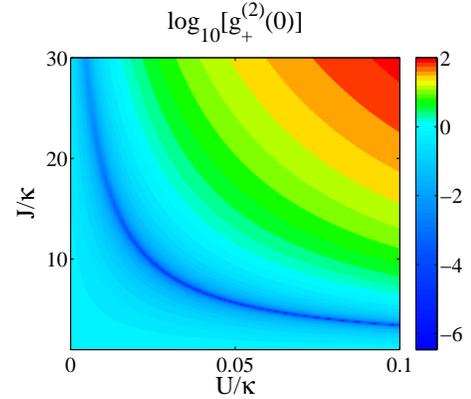}
\caption{(Color online) Logarithmic plot (of base $10$) of the equal-time
second-order correlation function $g_{+}^{\left( 2\right) }\left( 0\right)$
as a function of nonlinear interaction strength $U/\protect\kappa $ and
coupling strength $J/\protect\kappa$ for $\Delta=-J$ and $\varepsilon=0.01\kappa$.}
\label{fig5}
\end{figure}

As the statistic properties of photons for the symmetric and antisymmetric
modes are similar to each other, let us focus on the case of
the symmetric mode in the following. A two-dimensional plot of
the equal-time second-order correlation function $g_{+}^{\left( 2\right)
}\left( 0\right)$ as a function of nonlinear interaction strength $U/\kappa $
and coupling strength $J/\kappa$ for $\Delta =-J $ is shown in
Fig.~\ref{fig5}. With increasing $J/\kappa$, the value of $U/\kappa $ for
getting the strong antibunching (dark blue regime in the figure) descends gradually. That is to say, the
requirement of the nonlinear interaction strength $U$ for obtaining strong
photon antibunching in the symmetric mode can be controlled by tuning the value of the
coupling strength $J$ in the photonic molecule.

\begin{figure}[tbp]
\includegraphics[bb=18 15 372 276, width=6 cm, clip]{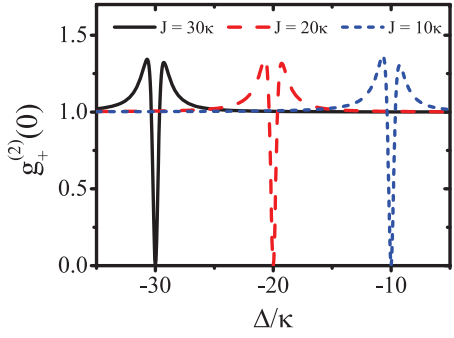}
\caption{(Color online) The equal-time second-order correlation function $%
g_{+}^{\left( 2\right)}\left( 0\right)$ as a function of the detuning $%
\Delta/\kappa$ for different values of the coupling strength $J$ with
nonlinear interaction strength $U/\protect\kappa=\protect\kappa/(4J)$ and Rabi frequency $\varepsilon=0.01\kappa$:
(black solid line) $J =30 \protect\kappa$; (red dash line) $J =20 \protect%
\kappa$; (blue short dash line) $J =10 \protect\kappa$.}
\label{fig6}
\end{figure}

The equal-time second-order correlation function $g_{+}^{\left( 2\right)
}\left( 0\right)$ as a function of the detuning $\Delta/\kappa $
for different coupling strengths $J/\kappa$ are shown in Fig.~\ref%
{fig6}, where $U/\kappa=\kappa/(4J)$. With the increase of $J$, the optimal detuning for strong antibunching in the symmetric
mode shifts as $\Delta =-J$. As a consequence, we can shift the optimal
value of the detuning for strong antibunching in the symmetric mode by
tuning the coupling strength $J$ in the photonic molecule. This is
significantly different from the result given in Refs.~\cite
{LiewPRL10,BambaPRA11}, where the optimal detuning is fixed at $\Delta/\kappa =\pm 1/(2\sqrt{3})$ in the strong coupling condition $J\gg \kappa$~\cite{BambaPRA11}.
As the coupling between the microtoroids can be controlled precisely in the experiments~\cite{GrudininPRL10,PengNP14a,PengNP14b}, the symmetric and antisymmetric modes in photonic molecules with weak nonlinearity can be used to generate tunable single-photon sources.

\begin{figure}[tbp]
\includegraphics[bb=25 1 386 285, width=4.2 cm, clip]{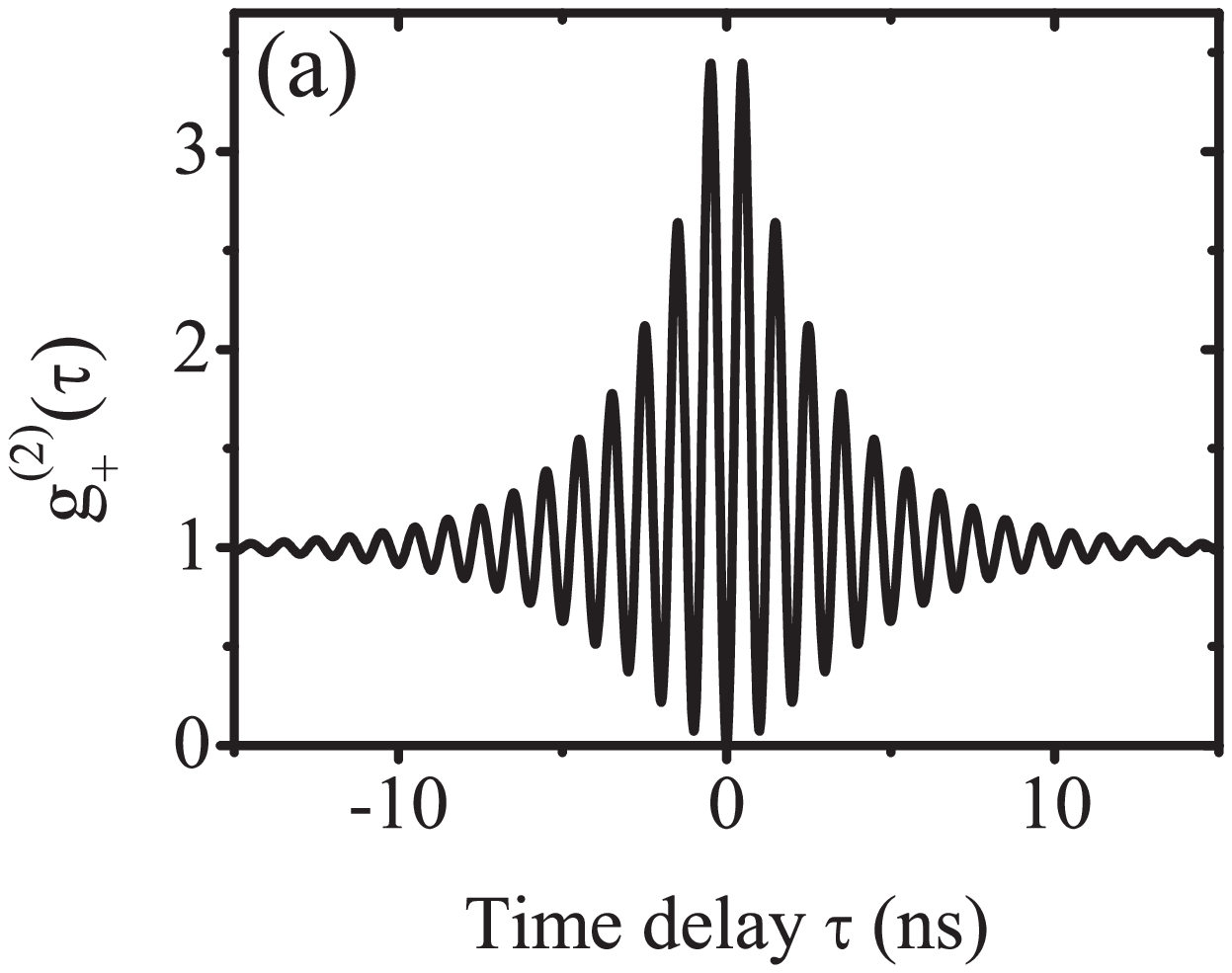}
\includegraphics[bb=25 1 386 285, width=4.2 cm, clip]{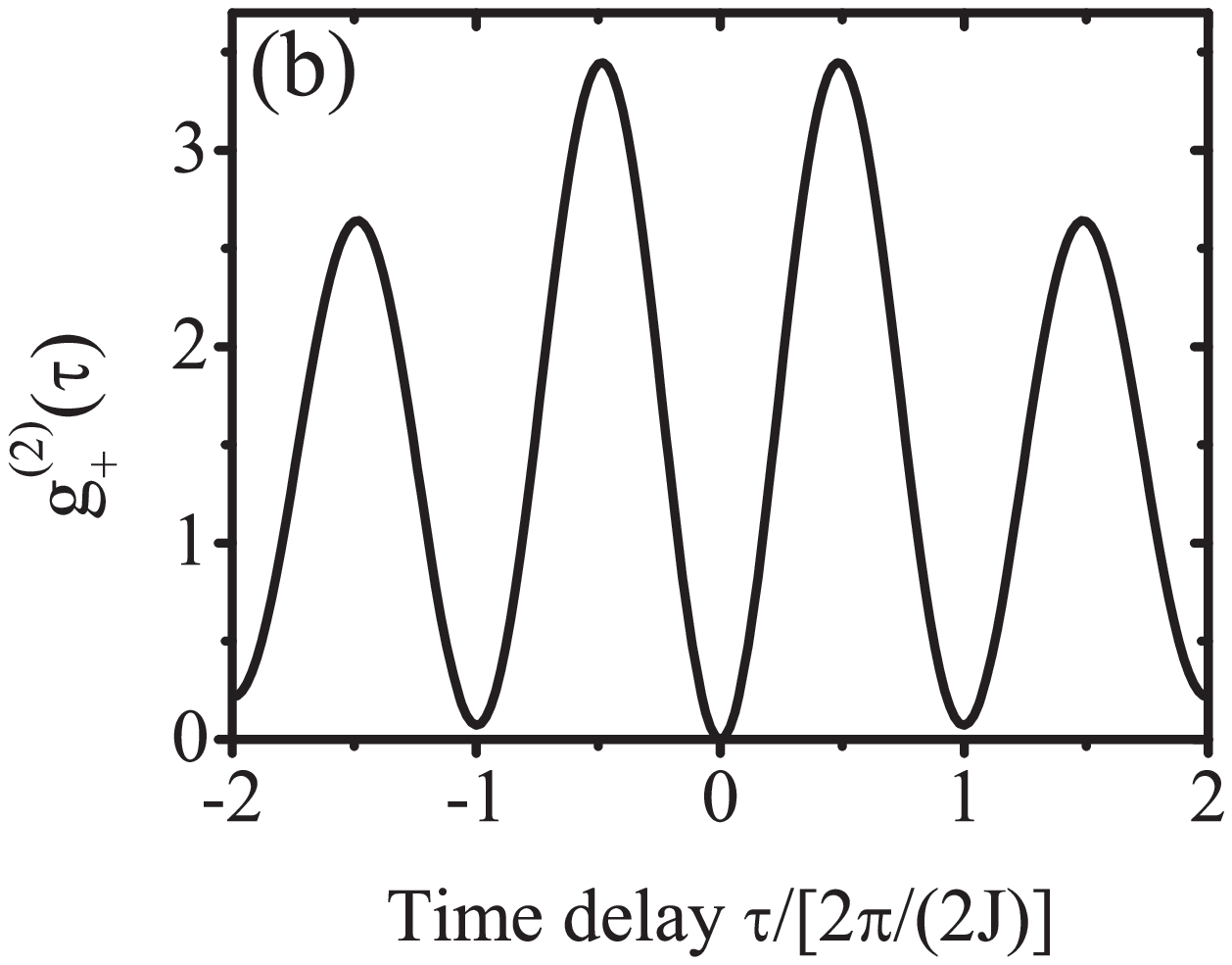}
\caption{(a) The second-order correlation functions $g_{+ }^{\left( 2\right) }\left( \tau \right)$ as a function of the time delay $\tau$. (b) $g_{+ }^{\left( 2\right) }\left( \tau \right)$ as a function of the normalized time delay $\tau/[2\pi/(2J)]$. The parameters are $\Delta=-5\kappa$, $J=5\kappa$, $U=0.05\kappa$, $\varepsilon=0.01\kappa$ and $\kappa=2\pi\times 100$ MHz.}
\label{fig7}
\end{figure}

Now, let us do some discussions about the feasibility of the strong photon antibunching effect for weak nonlinearity with some realistic parameters. For the experiment in Ref.~\cite{GrudininPRL10}, the resonance frequency for cavity mode is about $200$ THz, the Q-factor is $4 \times 10^7$ for empty cavity and the coupling strength between the two cavity modes ranges from $5$ MHz to nearly $5$ GHz. The Q-factor for the microtoroid made from silica doped with Kerr medium should become lower, and $Q=3 \times 10^6$ for the microtoroid made from silica doped with gain-medium was obtained in Ref.~\cite{PengNP14a}.

After considering the parameters in the experiments~\cite{GrudininPRL10,PengNP14a}, we take the parameters $\kappa=2\pi\times 100$ MHz and $J=5\kappa$ for numerical calculations, and the second-order correlation function $g_{+}^{\left( 2\right) }\left( \tau \right)$ is plotted as a function of the time delay $\tau$ in Fig.~\ref{fig7}. Similar to the reports given in Refs.~\cite{LiewPRL10,BambaPRA11,XuJPB13}, $g_{+}^{\left( 2\right) }\left( \tau \right)$ shows an oscillation behavior as the delay time going on with the period $2\pi/(2J)$ [as shown in Fig.~\ref{fig7}(b)]. The magnitude of the oscillation decreases as the increase of $\tau$ and almost approaches unity as $\tau \geq 10$ ns [as shown in Fig.~\ref{fig7}(a)], which is about the lifetime of the photons in the cavities. This oscillation behavior comes from the Rabi oscillation between the photon states and we will explain this in detail in the end of next section.

\section{Optimal conditions}

In order to understand the origin of the above strong antibunching obtained numerically, we will derive
the optimal conditions analytically following the method given in Ref.~\cite{BambaPRA11}. In the weak driving condition $%
\varepsilon \ll \kappa $, we expand the wave function on a Fock-state
basis of symmetric and antisymmetric modes truncated to the two-photon
manifold with the ansatz:
\begin{eqnarray}
\left\vert \psi \right\rangle  &=&C_{00}\left\vert 0,0\right\rangle
+C_{10}\left\vert 1,0\right\rangle +C_{01}\left\vert 0,1\right\rangle
\notag  \label{eq:5} \\
&&+C_{20}\left\vert 2,0\right\rangle +C_{11}\left\vert 1,1\right\rangle
+C_{02}\left\vert 0,2\right\rangle .
\end{eqnarray}%
Here, $\left\vert n_{+},n_{-}\right\rangle $ represents the Fock state with $%
n_{+}$ photons in the symmetric mode and $n_{-}$ photons in the
antisymmetric mode. Substituting the wave function [Eq.~(\ref{eq:5})] and
Hamiltonian [Eq.~(\ref{eq:2})] into the Schrodinger's equation, we get the
dynamic equations for the coefficients $C_{n_{+}n_{-}}$:
\begin{eqnarray}
i\frac{\partial }{\partial t}C_{00} &=&\frac{\varepsilon }{\sqrt{2}}C_{10}+%
\frac{\varepsilon }{\sqrt{2}}C_{01},  \nonumber \\
i\frac{\partial }{\partial t}C_{10} &=&\left( \Delta -J-i\frac{\kappa }{2%
}\right) C_{10}+\frac{\varepsilon }{\sqrt{2}}\left( C_{00}+C_{11}\right)
+\varepsilon C_{20},  \nonumber \\
i\frac{\partial }{\partial t}C_{01} &=&\left( \Delta +J-i\frac{\kappa }{2%
}\right) C_{01}+\frac{\varepsilon }{\sqrt{2}}\left( C_{00}+C_{11}\right)
+\varepsilon C_{02},  \nonumber \\
i\frac{\partial }{\partial t}C_{20} &=&\left[ U+2\left( \Delta -J-i\frac{%
\kappa }{2}\right) \right] C_{20}+\varepsilon C_{10}+UC_{02}, \nonumber
\\
i\frac{\partial }{\partial t}C_{02} &=&\left[ U+2\left( \Delta +J-i\frac{%
\kappa }{2}\right) \right] C_{02}+\varepsilon C_{01}+UC_{20},  \nonumber
\\
i\frac{\partial }{\partial t}C_{11} &=&\frac{\varepsilon }{\sqrt{2}}\left(
C_{01}+C_{10}\right) +\left( 2\Delta -i\kappa +2U\right) C_{11}.
\label{eq:11}
\end{eqnarray}

Under the weak driving condition $\varepsilon \ll \kappa $, we have $|C_{00}|\gg |C_{10}|,|C_{01}|\gg |C_{20}|,|C_{11}|,|C_{02}|$, and the equations for the coefficients of one-photon states,
\begin{eqnarray}
\left( \Delta-J-i\frac{\kappa }{2}\right) C_{10} &=&-\frac{\varepsilon
}{\sqrt{2}}C_{00},  \label{eq:12} \\
\left( \Delta+J-i\frac{\kappa }{2}\right) C_{01} &=&-\frac{\varepsilon
}{\sqrt{2}}C_{00},  \label{eq:13}
\end{eqnarray}%
and for the coefficients of two-photon states,
\begin{eqnarray}
0 &=& \left[ U+2\left( \Delta -J-i\frac{\kappa }{2}\right) \right]
C_{20}+UC_{02}+\varepsilon C_{10},  \label{eq:14} \\
0 &=& \left[ U+2\left( \Delta +J-i\frac{\kappa }{2}\right) \right]
C_{02}+UC_{20}+\varepsilon C_{01},  \label{eq:15} \\
0 &=& \left( 2\Delta -i\kappa +2U\right) C_{11}+\frac{\varepsilon }{\sqrt{2}}%
\left( C_{01}+C_{10}\right).  \label{eq:16}
\end{eqnarray}%
From Eqs.~(\ref{eq:12})-(\ref{eq:13}), the relation between $C_{10}$\ and $%
C_{01}$ reads
\begin{equation}  \label{eq:17}
\frac{C_{10}}{C_{01}}=\frac{ \Delta +J-i\frac{\kappa }{2} }{%
\Delta-J-i\frac{\kappa }{2}}=\frac{1}{\eta}.
\end{equation}%
Substituting this relation into Eqs.~(\ref{eq:15})-(\ref{eq:16}), we get
\begin{eqnarray}
0 &=&\left[ U+2\left( \Delta +J-i\frac{\kappa }{2}\right) \right]
C_{02}+UC_{20}+\varepsilon \eta C_{10},  \label{eq:19} \\
0 &=&\left( 2\Delta -i\kappa +2U\right) C_{11}+\frac{ \varepsilon }{
\sqrt{2}}\frac{2\Delta -i\kappa }{\Delta -J-i\frac{\kappa }{2}}
C_{01}.  \label{eq:20}
\end{eqnarray}

The conditions for $g_{+}^{\left( 2\right)}\left( 0\right) \ll 1$ are derived from Eqs.~(\ref{eq:14}) and (\ref{eq:19}) by setting $C_{20}=0$, so we get%
\begin{eqnarray}
0 &=&UC_{02}+\varepsilon C_{10},  \label{eq:21} \\
0 &=&\left[ U+2\left( \Delta +J-i\frac{\kappa }{2}\right) \right]
C_{02}+\varepsilon \eta C_{10}.  \label{eq:22}
\end{eqnarray}%
The condition for that $C_{10}$ and $C_{02}$ have non-trival solutions is that the
determinant of the coefficient matrices of Eqs.~(\ref{eq:21})-(\ref{eq:22})
equals to zero, then we get the equation for optimal photon antibunching as
\begin{equation}  \label{eq:23}
\frac{\kappa ^{2}}{4}-JU-\left( \Delta +J\right) ^{2}+i\left( \Delta+J\right) \kappa =0.
\end{equation}%
For the imagine part to be zero, we have
\begin{equation}  \label{eq:24}
\Delta _{\mathrm{opt}}=-J,
\end{equation}%
and for the real part equal to be zero, we get
\begin{equation}  \label{eq:25}
\frac{U_{\mathrm{opt}}}{\kappa }=\frac{\kappa }{4J}.
\end{equation}%
Eqs.~(\ref{eq:24})-(\ref{eq:25}) are the optimal conditions for $g_{+}^{\left( 2\right)}\left( 0\right) \ll 1$ in Fig.~\ref{fig2}-\ref{fig5}.

Similarly, we can get the optimal conditions for $g_{-}^{\left( 2\right)}\left( 0\right) \ll 1$ from Eqs.~(\ref{eq:14})-(\ref{eq:16}) by using the relation between $C_{10}$\ and $C_{01}$ [Eq.~(\ref{eq:17})] and setting $C_{02}=0$, then we get
\begin{eqnarray}
0 &=&\left[ U+2\left( \Delta -J-i\frac{\kappa }{2}\right) \right]
C_{20}+ \frac{\varepsilon}{\eta}C_{01},  \label{eq:26} \\
0 &=&UC_{20}+\varepsilon C_{01}.  \label{eq:27}
\end{eqnarray}%
To make sure $C_{01}$ and $C_{20}$ have non-trival solutions, we have
\begin{equation}  \label{eq:28}
\frac{\kappa ^{2}}{4}+JU-\left( \Delta -J\right) ^{2}+i\left( \Delta-J\right) \kappa =0.
\end{equation}%
Then, the optimal conditions for $g_{-}^{\left( 2\right)}\left( 0\right) \ll 1$ are given by
\begin{eqnarray}  \label{eq:29}
\Delta _{\mathrm{opt}} &=&J, \\
\frac{U_{\mathrm{opt}}}{\kappa } &=&-\frac{\kappa }{4J}.
\end{eqnarray}

\begin{figure}[tbp]
\includegraphics[bb=70 133 518 571, width=8 cm, clip]{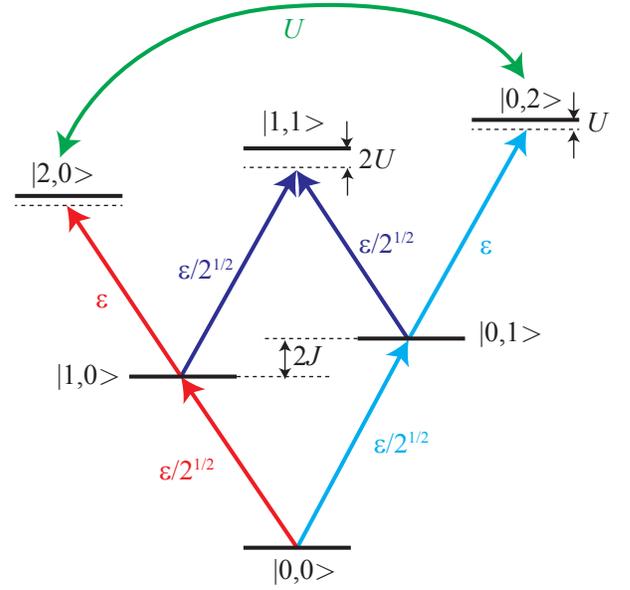}
\caption{(Color online) Energy-level diagram showing the zero-, one- and
two-photon states (horizontal black short lines) and the transition paths
leading to the quantum interference responsible for the strong antibunching
(color lines with arrows). $\left\vert n_{+},n_{-}\right\rangle$ represents
the Fock state with $n_{+}$ photons in the symmetric mode and $n_{-}$
photons in the antisymmetric mode. }
\label{fig8}
\end{figure}

In Fig.~\ref{fig8}, we show the energy-level diagram and the transition
paths. There are two paths for two-photon excitation in the symmetric mode:
(i) directly exciting two photons in the symmetric mode (red lines with
arrows), i.e. $\left\vert 0,0\right\rangle \overset{\varepsilon /\sqrt{2}}{
\rightarrow }\left\vert 1,0\right\rangle \overset{\varepsilon }{\rightarrow }
\left\vert 2,0\right\rangle $; (ii) exciting two photons in the
antisymmetric mode (cyan lines with arrows), then coupling to the the
symmetric mode via the nonlinear interaction (green line with arrows): i.e. $%
\left\vert 0,0\right\rangle \overset{\varepsilon /\sqrt{2}}{\rightarrow }
\left\vert 0,1\right\rangle \overset{\varepsilon }{\rightarrow }\left\vert
0,2\right\rangle \overset{U}{\rightarrow }\left\vert 2,0\right\rangle $.
These two paths lead to the destructive quantum interference that is
responsible for the strong antibunching in the symmetric mode. Moreover, the fact that
the optimal nonlinear interaction strength is inversely proportional to the
coupling strength $J$ [$U_{\mathrm{opt}}/\kappa=\kappa /(4J)$] can be understood as follows: For $\Delta _{\mathrm{opt}}=-J$, with the increase of
the coupling strength between the cavity modes $J$, the non-resonant
transition $\left\vert 0,0\right\rangle \overset{\varepsilon /\sqrt{2}}{
\rightarrow }\left\vert 1,0\right\rangle \overset{\varepsilon }{\rightarrow }
\left\vert 2,0\right\rangle $ will be suppressed, while the resonant
transition $\left\vert 0,0\right\rangle \overset{\varepsilon /\sqrt{2}}{
\rightarrow }\left\vert 0,1\right\rangle \overset{\varepsilon }{\rightarrow }
\left\vert 0,2\right\rangle $ will be enhanced, so the nonlinear interaction
strength $U$ needed for destructive quantum interference becomes smaller.
Similar origin leads to the strong antibunching effect in the antisymmetric
mode.

Finally, let us give an explanation for the oscillation behavior of $g_{+}^{\left( 2\right) }\left( \tau \right)$ (as shown in Fig.~\ref{fig7}) via the energy-level diagram (Fig.~\ref{fig8}). In the short time approximation $t \ll 2\pi/\kappa \ll 2\pi/\varepsilon$, we can treat the transitions $\left\vert 0,0\right\rangle \overset{\varepsilon /\sqrt{2}}{
\rightarrow }\left\vert 1,0\right\rangle$ and $\left\vert 0,0\right\rangle \overset{\varepsilon /\sqrt{2}}{
\rightarrow }\left\vert 0,1\right\rangle$ as two individual Rabi models with the system in the vacuum state initially. For $\Delta=-J$, the driving field is resonant with the transitions $\left\vert 0,0\right\rangle
\rightarrow \left\vert 0,1\right\rangle$ while the detuning between driving field and the transitions $\left\vert 0,0\right\rangle \rightarrow \left\vert 1,0\right\rangle$ is $2J$, so we have $|C_{01}|^{2} \simeq \left[1-\cos(\sqrt{2}\epsilon t)\right]/2$ and $|C_{10}|^{2} \simeq [1-\cos(2Jt)]\varepsilon^2/(4J^2)$. The time oscillation of $g_{+}^{\left( 2\right) }\left( \tau \right)$ with period $2\pi/(2J)$ comes from the Rabi oscillation between $|0,0\rangle$ and $|1,0\rangle$.

\section{Conclusions}

In summary, we have studied the photon statistics of the symmetric and
antisymmetric modes in the photonic molecule consisting of two linearly coupled
nonlinear cavity modes. Due to the destructive quantum interference effect
between the different paths for two-photon excitation, the photons of both the
symmetric and antisymmetric modes can exhibit strong antibunching effect
even with weak nonlinear interaction in the photonic molecule. By analytical
method, we show that the optimal frequency detunings for strong photon antibunching in the
symmetric and antisymmetric modes are linearly dependent on the
coupling strength between the nonlinear cavity modes in the photonic molecule. Thus
we can control the statistic properties of the photons by tuning the
coupling strength between the nonlinear cavity modes in the photonic molecule. Our
results may have important applications in generating tunable single-photon
sources.

\vskip 2pc \leftline{\bf Acknowledgement}

We thank Y. L. Liu, L. Ge, X. Xiao, Q. Zheng and Y. Yao for fruitful discussions.
This work is supported by the Postdoctoral Science Foundation of
China (under Grant No. 2014M550019), the NSFC (under Grant No. 11174027),
and the National 973 program (under Grant No. 2012CB922104 and No.
2014CB921402).

\bibliographystyle{apsrev}
\bibliography{ref}

\end{document}